\documentclass[12pt]{iopart}
\usepackage[]{epsfig}
\usepackage{graphicx}
\begin{document}

\title{Energy landscape - a key concept in the dynamics of liquids and glasses}

\author{U. Buchenau
\footnote[3]{u.buchenau@fz-juelich.de} }
\address {Institut f\"ur Festk\"orperforschung, Forschungszentrum
        J\"ulich, Postfach 1913, 52425 J\"ulich, Germany}


\begin{abstract}
There is a growing belief that the mode coupling theory is the
proper microscopic theory for the dynamics of the undercooled
liquid above and around a critical temperature $T_c$. In addition,
there is some evidence that the system leaves the saddlepoints of
the energy landscape to settle in the valleys at this critical
temperature. Finally, there is a microscopic theory for the
entropy well below $T_c$ (i.e. close to the calorimetric glass
transition $T_g$) by M$\acute{e}$zard and Parisi, which allows to
calculate the Kauzmann temperature from the atomic pair
potentials.

The dynamics of the frozen glass phase is at present limited to
phenomenological models. In the spirit of the energy landscape
concept, one considers an ensemble of independent asymmetric
double-well potentials with a wide distribution of barrier heights
and asymmetries (ADWP or Gilroy-Phillips model). The model gives
an excellent description of the relaxation of glasses up to about
$T_g/4$. Above this temperature, the interaction between different
relaxation centers begins to play a role. In a mean-field
treatment, the interaction reduces the number of relaxation
centers needed to bring the shear modulus down to zero by a factor
of three.
\end{abstract}

\maketitle

\section{Introduction}

In the authors view, our present foggy picture of the glass
transition begins to show some cornerstones of a future solid
theoretical building, namely the energy landscape concept
(Goldstein 1968; Johari and Goldstein 1970, 1971; Stillinger
1995), the mode coupling theory (Bengtzelius \etal 1984; G\"otze
and Sj\"ogren 1992) together with the realization (Bengtzelius
\etal 1984; Angell 1988) that the critical temperature of this
theory marks the onset of thermally activated motion between the
minima of the energy landscape, and, finally, the calculation of
the Kauzmann temperature from the interatomic potentials
(M$\acute{e}$zard and Parisi 1996, 1999). These theories explain,
at least in principle, the dynamics above and around $T_c$ as well
as the thermodynamics near to the calorimetric glass transition
$T_g$, leaving only the fragility (Angell 1988; B\"ohmer \etal
1993; Angell 1995) without a solid theoretical foundation.

The present paper expands this view in a bit more detail in the
next two sections. Section 4 addresses the relaxation in glasses
at temperatures well below $T_g$ in terms of thermally activated
jumps in an ensemble of independent asymmetric double-well
potentials (the ADWP (\underline{A}symmetric \underline{D}ouble
\underline{W}ell \underline{P}otential) (Pollak and Pike 1972) or
Gilroy-Phillips model (Gilroy and Phillips 1981)). Section 5
considers the effect of the interaction between different
relaxation centers, which is shown to become dominant at the glass
transition. Section 6 compares the prediction of a mean field
treatment to experimental data on the breakdown of the shear
modulus. Conclusions are given in Section 7.

\section{$T_c$: The onset of thermally activated motion}

\begin{figure}[b]
\begin{center}
\hspace{1.7cm}
 \epsfig{file=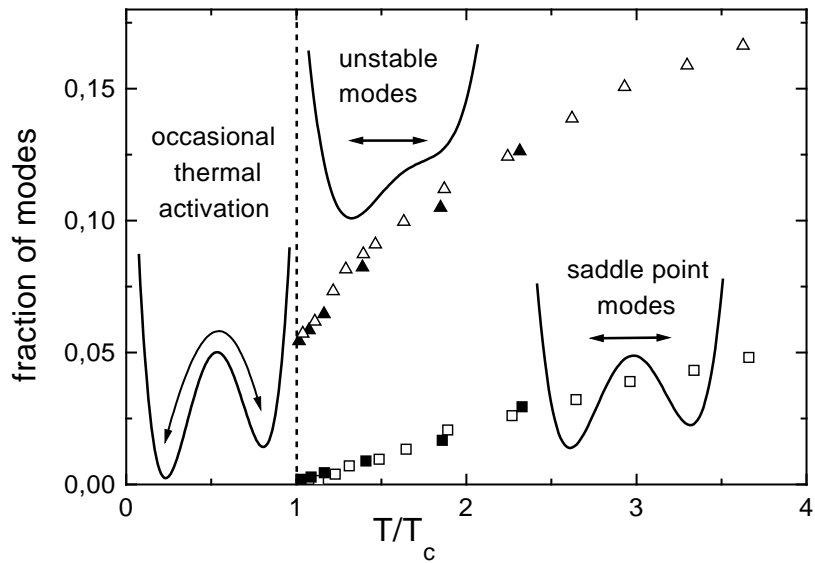,height=9cm,angle=0}
%

\end{center}
\caption{Temperature dependence of the fraction of saddle-point
modes (squares) and of instantaneous unstable normal modes
(triangles) in two Lennard-Jones glass formers after Angelani
\etal (2000).}

\end{figure}

The mode coupling theory of the glass transition (G\"otze and
Sj\"ogren 1992) does not require the concept of the energy
landscape. In fact, its most impressive experimental proof was
found in colloids (Pusey and van Megen 1986; van Megen and
Underwood 1994), which do not have an energy landscape.

On the other hand, the dynamics of the undercooled liquid at lower
temperatures is dominated by thermally activated hopping between
different valleys of the energy landscape. The conjecture
(Bengtzelius \etal 1984; Angell 1988) of an onset of this
thermally activated motion at the critical temperature $T_c$ of
the mode coupling theory has found more and more support from
numerical simulations of model glass formers (Schr\o der \etal
2000; Angelani \etal 2000; Broderix \etal 2000; but see also
Doliwa and Heuer 2002, which stress the important role of thermal
activation above $T_c$). The main result is illustrated in Fig. 1,
adapted from Angelani \etal (2000). As one lowers the temperature
towards $T_c$, the average number of saddle points of the energy
landscape on which the system finds itself at a given moment in
time decreases. This number extrapolates to zero at $T_c$. At
$T_c$, one still finds a finite number of unstable instantaneous
normal modes. However, these stem from shoulders of the potential
with a negative curvature rather than from true saddle points.
They give rise to the fast picosecond motion (Angell 1995), but
they do not dominate the long time dynamics.

Above $T_c$, the separation of the $\alpha$-process (the
elementary process of the flow) from the microscopic picosecond
motion (Franosch \etal 1998) in an undercooled liquid seems to be
reasonably well described by the mode coupling theory. This was
demonstrated by neutron (Knaak \etal 1988; Frick \etal 1991;
Wuttke \etal 1993) and light scattering experiments (Li \etal
1992; Sokolov 1998; Wiedersich \etal 2000a) on ionic, molecular
and polymeric glass formers, as well as in a number of numerical
simulations (Kob and Andersen 1995a, 1995b; Nauroth and Kob 1997;
Kammerer \etal 1998a, 1998b). There are more examples (G\"otze
1999). One finds the proper scaling relations for the time and
temperature dependence of the $\alpha$-process and the fast
picosecond $\beta$ process, consistent with the exponents
determined from the temperature dependence of the viscosity above
$T_c$. As a general rule (Sokolov 1998), one finds
$\tau_\alpha(T_c)\approx 10^{-7}$ s, a bit longer than the
originally considered value (Goldstein 1968; Angell 1988) of
$10^{-9}$ s.

Note this does not imply a perfect fit of theory and experimental
data. Though one observes the expected power law behaviour of the
viscosity in many liquids (Taborek \etal 1986), accurate
dielectric data in salol show a temperature dependence of the
power law exponent of $\tau_\alpha$ even well above $T_c$ (Stickel
\etal 1995). Quantitative checks of the theory are at present
impeded by the difficulty to calculate $T_c$ and the exponents for
most real glass formers.

Nevertheless, the mode coupling theory is a true microscopic
theory, which in principle allows to calculate its parameters from
a knowledge of the interatomic potentials. The results obtained so
far seem to show that one can identify the critical temperature of
this microscopic theory with the temperature at which the system
leaves the saddle points of the energy landscape to settle in its
valleys. Below $T_c$, one expects a gradual transition to
thermally activated energy landscape dynamics. Though there is no
generally agreed description of this landscape dynamics, there is
at least a microscopic theory (M$\acute{e}$zard and Parisi 1996,
1999) for the thermodynamics of the undercooled liquid at this
lower temperature.

\section{Calculating the Kauzmann temperature}

This further important theoretical progress of the last decade
concerns an old concept from the thermodynamics of the glass
transition, the Kauzmann temperature. The concept stems from the
experimental observation that the entropy difference between
undercooled liquid and crystal seems to extrapolate to zero at a
finite temperature, the Kauzmann temperature $T_K$, at which the
glass former in principle condenses into a single structural
configuration. Since the viscosity depends on the number of
accessible configurations, one expects a divergence of the
viscosity at about the same temperature (an excellent review of
the older empirical attempts to model the thermodynamics and
kinetics of the undercooled liquid has been given by J\"ackle
(1986)). In this general sense, the Kauzmann temperature is not
only important for the thermodynamics, but also for the dynamics
close to $T_g$. In fact, the empirical Adam-Gibbs model identifies
the temperature $T_K$ with the Vogel-Fulcher temperature $T_0$ of
the empirical VFT (Vogel-Fulcher-Tammann) or WLF
(Williams-Landel-Ferry (Ferry 1980)) relation
\begin{equation} \tau_\alpha=\tau_0{\rm e}^{A/(T-T_0)},
\label{vogel}
\end{equation}
where $\tau_0$ is a microscopic time and $A$ is a second parameter
of this empirical relation. If one looks more closely (Stickel
\etal 1995, 1996; Hansen \etal 1997), the Vogel-Fulcher relation
does not describe the temperature dependence of $\tau_\alpha$ very
well. From this data collection, one rather feels that each glass
former behaves differently below $T_c$. Nevertheless, the general
tendency of a divergence of the viscosity as the glass former
looses its configurational entropy cannot be denied.

M$\acute{e}$zard and Parisi (1996, 1999) have devised a recipe to
calculate the entropy and the Kauzmann temperature from the pair
potentials of a given glass former, thus providing the Kauzmann
extrapolation scheme with a theoretical solidity which it lacked
before. The calculation assumes an undercooled liquid below $T_c$
which spends most of its time vibrating in a local minimum of the
free energy, with only occasional jumps into a neighboring
minimum. In this situation, one assumes the validity of the
harmonic approximation for the motion within the single well.
Using the replica concept, one calculates the free energy as a
function of temperature and finds a nonzero Kauzmann temperature.
This is again a microscopic theory, because it allows to calculate
the heat capacity and the Kauzmann temperature from the pair
potentials between the atoms. Though the Kauzmann temperature
itself might still prove to be an artefact of the mean-field
approximation of the calculation, the calculated heat capacity
above this temperature resembles measured data above $T_g$ both in
shape and size.

The next section proceeds to a phenomenological model of the
thermally activated energy landscape dynamics in the
low-temperature glass phase.

\section{ADWP or Gilroy-Phillips model}

The ADWP (Asymmetric-Double-Well-Potential) (Pollak and Pike 1972)
or Gilroy-Phillips model (Gilroy and Phillips 1981) is a member of
a family of three glass models, which are essentially one and the
same model applied to three different situations (the other two
are the tunneling model (Phillips 1981) and the soft-potential
model (Parshin 1994)). The basic idea is to simplify the
multiminimum situation of the energy landscape to an ensemble of
independent double-well potentials for local structural
rearrangements with a broad distribution of different barrier
heights $V$ and asymmetries $\Delta$ between the two minima (see
Fig. 2) - an inherently heterogeneous description of the dynamics
of the glass phase (for a review on the heterogeneity of
undercooled liquids see Richert 2002).

\begin{figure}[b]
\hspace{2cm} \vspace{0cm} \epsfig{file=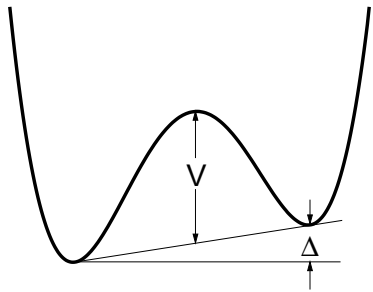,width=6
cm,angle=-90} \vspace{0cm} \caption{Asymmetric double-well
potential.}
\end{figure}

Consider a single relaxing entity, i.e. a single barrier of height
$V$ separating two neighboring energy minima. Fig. 2 shows
schematically the energy as a function of the configurational
coordinate going from one minimum to the other. In numerical
simulations of model glasses, one finds that this configurational
coordinate involves the motion of about five to fifty atoms in the
center of the relaxing entity (Heuer and Silbey 1996; Schober and
Oligschleger 1996) . There is no reason why the two minima should
have the same energy, so there will be an energy difference
$\Delta$ between them. A further characteristic of the two
adjacent minima is the coupling of this relaxing entity to the
external shear strain $\epsilon$. This is given by the coupling
constant $\gamma$, defined such that the asymmetry changes from
$\Delta$ to $\Delta+\gamma\epsilon$ under the applied shear strain
$\epsilon$.

With these three energies, the barrier height $V$, the asymmetry
$\Delta$ and the coupling constant $\gamma$, one can quantify the
contribution of this single relaxing entity to the dynamical
mechanical behaviour of the viscoelastic medium. The relaxation
time $\tau_V$ is given by the Arrhenius relation
\begin{equation} \tau_V=\tau_0{\rm e}^{V/k_BT},
\label{arrh}
\end{equation}
where $\tau_0$ is a microscopic time of the order of 10$^{-13}$
seconds, $V$ is the energy of the barrier between two energy
minima of the system, and $T$ is the temperature.

In the simplest possible approximation, the free energy $F$ of the
single relaxation center reads
\begin{equation}\label{free}
F =-k_BT\ln\left[2\cosh\left(\frac{\Delta+\gamma\epsilon}{2k_BT}
\right)\right].
\end{equation}
Its second derivative with respect to the shear distortion
$\epsilon$ is
\begin{equation}\label{asymm}
\frac{\partial^2F}{\partial\epsilon^2}
 =-\frac{\gamma^2}{4k_BT\cosh^2(\Delta/2k_BT)}
\end{equation}
The second derivative determines the contribution of the specific
relaxing entity to the difference between the shear moduli at
infinite and zero frequency. The main influence on the shear
modulus is due to relaxation in potentials with asymmetries
smaller than $k_BT$; for larger asymmetries the influence
decreases rapidly because of the square of the hyperbolic cosine
in the denominator.

We assume a number density function $n(V,\Delta)$ of these local
structural relaxations, which varies little if $V$ or $\Delta$
change by energies of the order $k_BT$. Integrating over the
asymmetry (Buchenau 2001), we obtain a net difference $\delta G$
between the infinite and the zero frequency shear moduli due to
relaxations with barrier heights between $V$ and $V+\delta V$
\begin{equation}\label{f0}
\frac{\delta G}{G\delta V}=\frac{\gamma^2n(V,0)}{G}\equiv f_0(V),
\end{equation}
where $G$ is the infinite frequency shear modulus. This relation
defines $f_0(V)$, the barrier density function {\it without}
interaction.

In the frozen glass phase, one expects a frozen-in distribution of
barrier heights and asymmetries and thus a temperature-independent
$f_0(V)$. One can check that by looking at the distribution with
low and high frequencies; according to the Arrhenius relation, eq.
(\ref{arrh}), one should be able to observe the same barrier
height at different temperatures. Such checks have been done for a
number of different glass formers. As long as one stays at
temperatures much lower than the glass temperature, one finds
impressive agreement with the idea of a temperature-independent
barrier density function. One example is vitreous silica up to 300
K, glass temperature 1473 K (Wiedersich \etal 2000b). Another one
is polymethylmethacrylate (PMMA)  below 80 K, glass temperature
383 K (Buchenau \etal 2002). For higher temperatures, however, the
barrier density function tends to increase with increasing
temperature (Surovtsev \etal 1998, Caliskan \etal 2002).

As will be shown in the next section, one has to expect such an
increase, because the interaction between different relaxation
centers determines the dynamics close to the breakdown of the
shear modulus. In this view, the Johari-Goldstein $\beta$-process,
a broad relaxation maximum close to the $\alpha$-process which
shows an Arrhenius behaviour both below and above $T_g$ (Johari
and Goldstein 1970, 1971; Kudlik \etal 1999) corresponds to a peak
in $f_0(V)$, strongly enhanced by the proximity of the breakdown.

\section{The 1/3-rule}

For the purpose of this and the following section, let us assume
that the decomposition of the complex energy landscape into an
ensemble of single relaxation centers or single relaxing entities
is a reasonable and solid basis. In a mean-field scheme, the
interaction between the single entity and all the others is taken
into account by embedding the single center into the viscoelastic
medium, calculating the viscoelastic properties selfconsistently.

The first implication of this mean-field assumption is that the
asymmetry $\Delta$ is no longer fixed, but changes on the Maxwell
time scale, because the viscoelastic medium is free to flow.
Consequently, the term $\gamma\epsilon$ in eq. (\ref{free}) can
adapt on the Maxwell time scale, thus changing $\Delta$ to a
different value.  The Maxwell time $\tau_M$ is given by the shear
viscosity $\eta$ and the infinite frequency shear modulus G
\begin{equation}\label{maxwell}
\tau_M=\eta/G,
\end{equation}
where all three quantities depend on temperature, the infinite
frequency shear modulus $G$ only weakly, but the two other
quantities drastically.

A change in $\Delta$ does not mean that the energy landscape
itself flows; to take a three-dimensional example, the barrier is
like a ridge between two sloping valleys with different slope;
going along the ridge changes the height difference of the two
valleys. In this example, the coordinate along the ridge could
correspond to the external shear strain; then one has a continuous
change of $\Delta$. It could also correspond to the
configurational coordinate of another relaxing entity in the
neighborhood, which changes the local shear strain at the given
relaxation center in a discontinuous way.

It is interesting to consider the consequences of a freely
changing value of $\Delta$. Taking eq. (\ref{free}) literally, one
calculates a Boltzmann factor of the relaxing entity
\begin{equation}\label{boltzmann}
\exp(-F/k_BT) =2\cosh(\Delta/2k_BT),
\end{equation}
which has its lowest value at the symmetric case, $\Delta=0$, and
diverges with increasing $\Delta$. This latter feature is of
course unphysical, because one cannot expect to gain energy
without limit by increasing the asymmetry of such a local entity.
However, the consideration shows that one must expect a relatively
low probability for the symmetric case, because it is
energetically unfavorable. In fact, both the number of tunneling
states and the excess entropy of the frozen glass seem to decrease
in selenium upon aging (Johari 1986).

As soon as $\Delta$ is able to change, the relaxing entity has an
additional possibility to find its thermal equilibrium, namely by
lowering the energy of the minimum in which the system happens to
find itself. It is therefore natural to assume that relaxation
centers with high barriers, whose relaxation time exceeds the
Maxwell time, do not contribute to the viscoelastic properties of
the medium, while those with shorter relaxation times have time to
equilibrate by jumps over the barrier and do contribute. The two
barrier regimes are separated by the Maxwell barrier $V_M$ with
\begin{equation}\label{vm}
V_M=k_BT\ln(\tau_M/\tau_0).
\end{equation}

Consider the energetics of a single double-well, with a barrier
low enough to equilibrate within the Maxwell time. Suppose a small
constant shear strain $\epsilon$ is switched on at time zero, with
the population of the minima of the double-well in thermal
equilibrium with respect to zero shear strain. The new thermal
equilibrium requires a number of jumps
\begin{equation}\label{nj}
\delta n=\frac{\gamma\epsilon}{4k_BT\cosh^2(\Delta/2k_BT)}.
\end{equation}
In order to calculate the energy $\delta U$ carried to the heat
bath, we have to multiply the number of jumps $\delta n$ with the
energy difference $\Delta+\gamma\epsilon$. Therefore these jumps
transport the energy
\begin{equation}\label{dU}
\delta
U=\frac{\Delta\gamma\epsilon+\gamma^2\epsilon^2}{4k_BT\cosh^2(\Delta/2k_BT)}
\end{equation}
from the macroscopic shear stress energy to the heat bath. The
term on the right hand side linear in $\epsilon$ must be
compensated by other relaxing entities with opposite sign of
$\Delta$ (otherwise there would be no initial equilibrium). If one
compares the second quadratic term $\delta U_2$ with the second
derivative of the free energy in eq. (\ref{asymm}), one finds that
it is twice as high as the free energy decrease $\delta F$
calculated from eq. (\ref{asymm})
\begin{equation}\label{deltaf}
\delta
F=\frac{\gamma^2\epsilon^2}{8k_BT\cosh^2(\Delta/2k_BT)}=\frac{1}{2}\delta
U_2,
\end{equation}
which determines the reduction of the shear modulus by
the barrier. The physical reason for this is the reduced entropy;
spending the energy one has spanned an entropic spring. Thus the
reduction of the shear modulus is only half that expected from the
spent energy. This is in principle textbook knowledge for a Debye
relaxation, but is explained here again, because it is essential
for the understanding of the glass transition.

In a Gedankenexperiment, let the thermally equilibrated relaxation
center return from the actual asymmetry $\Delta+\gamma\epsilon$ in
the strained state to its original asymmetry $\Delta$, say by
appropriate jumps in the surroundings which change the local
strain at the center. In principle, this return would again
require the energy $\delta U$, to be taken again from the
macroscopic stress energy. However, this return occurs on the
Maxwell time scale, which is long compared to the relaxation time
of the relaxation center. Therefore the population of its two
minima adapts adiabatically. The number of backjumps is again the
same $\delta n$, but now the asymmetry reduces gradually from
$\Delta+\gamma\epsilon$ to $\Delta$ in the course of the process,
reducing the average energy per jump to $\Delta+\gamma\epsilon/2$.
This means one needs only the energy $\delta U_2-\delta F=\delta
F$ to return.

Now imagine this happens for all relaxation centers with barriers
lower than the Maxwell barrier of eq. (\ref{vm}) in the
undercooled liquid. Then one returns to an unstrained equilibrium
in the strained state. This means a full relaxation of the initial
stress, as in the true flow process characterized by the Maxwell
time. Naturally, our Gedankenexperiment is a rather improbable
realization of this process, because in the real process a given
relaxation center will almost never return to its initial
asymmetry, though the macroscopic stress relaxes back to zero.
However, this special realization allows to keep track of the
energy contributions.

In this cycle from the initial equilibrium to a new equilibrium in
the strained state, each barrier with $\tau_V$ smaller than
$\tau_M$ takes the energy $3\delta F$ from the stress energy,
$2\delta F$ in the initial equilibration and $\delta F$ on the
Maxwell time scale. The direct reduction of the shear modulus, the
one expected if there were no interaction between different
barriers, corresponds only to a single $\delta F$. The energy
$3\delta F$ is taken from the potential energy in the strain
field, reducing it to zero. One arrives at the conclusion that the
stress relaxation occurs when the noninteracting relaxation
centers reduce the shear modulus by one third of its infinite
frequency value. To put it differently, the interaction between
relaxing entities reduces the number needed to bring the long time
shear modulus down to zero by a factor of three.

In terms of the barrier density function $f_0(V)$, this means
\begin{equation}\label{norm}
\int_0^{V_M} f_0(V)dV=\frac{1}{3}.
\end{equation}

This is the 1/3-rule, which allows to calculate the Maxwell
barrier (and from the Maxwell barrier, the shear viscosity) for a
given barrier density function $f_0(V)$. The barrier density
function $f_0(V)$ in turn can be determined from measurements at
times shorter than the Maxwell time. To do this, one needs a
quantitative treatment of the interaction between different
relaxation centers, which is the topic of the next section.

The 1/3-rule provides a qualitative understanding of the
fragility: As the temperature increases, $f_0(V)$ is expected to
increase, because symmetric double-well potentials are
energetically unfavorable, as pointed out at the beginning of this
section. This implies that $V_M$ decreases with increasing
temperature, as one indeed observes in experiment. The problem is
to make this understanding quantitative. One could hope to use the
theoretical tools of M$\acute{e}$zard and Parisi (1996, 1999) to
achieve this end.

\section{The breakdown of the shear modulus}

The 1/3-rule, eq. (\ref{norm}), can be derived independently
(Buchenau 2002) from the assumption
\begin{equation}\label{f}
\frac{\delta G}{G\delta V}= f_0(V)\frac{G^2}{G(\tau_V)^2}{\rm
e}^{-\tau_V/\tau_M}\equiv f(V).
\end{equation}

Eq. (\ref{f}) is a generalization of the definition of the barrier
density function $f_0(V)$ {\it without} interaction, eq.
(\ref{f0}), to describe the enhancement of the effect of a single
relaxation center by the interaction, together with the cutoff at
the Maxwell time. The equation defines a barrier density function
$f(V)$ {\it with} interaction, which can then be used to calculate
the full shear response.

One can justify the quadratic enhancement factor assuming a
constant strain applied at time zero. The relaxation will tend to
equilibrate at the time $\tau_V$, when the square of the stress -
a measure of the remaining stress energy - is reduced by precisely
this factor,  while the number of jumps required for the
equilibration may be taken to be unchanged. If the number  of
jumps and the distortion remain unchanged, the energy and the free
energy contribution remain unchanged. This implies that the
reduction of the stress energy by the relaxation is also
unchanged. This in turn increases the apparent barrier density
function $f(V)$ by the square of $G/G(\tau_V)$. A physical
interpretation of this increase are lower-barrier jumps, taking
place in the neighborhood after a jump of the central entity.

\begin{figure}[b]
\hspace{3cm} \vspace{0cm}
\epsfig{file=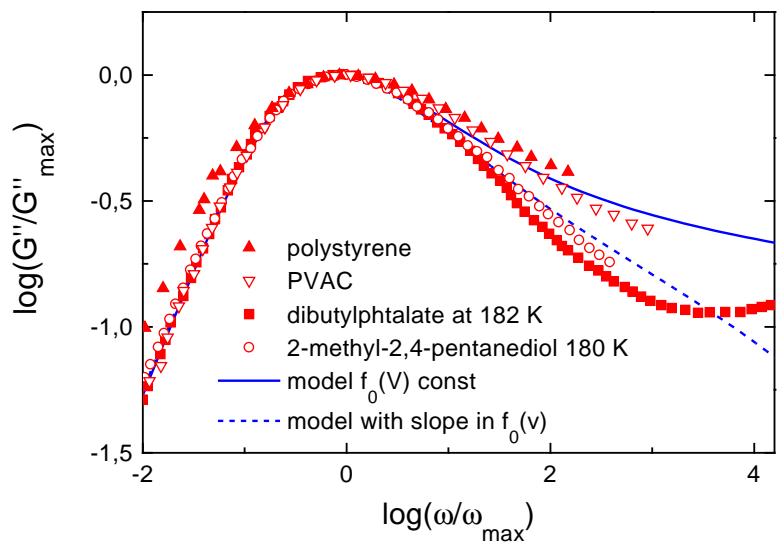,width=12cm,angle=0} \vspace{0cm}
\caption{Shape of the normalized $\alpha$-peak in $G''(\omega)$ in
a log-log plot for two polymers (Donth \etal 1996) and the two
molecular glass formers dibutylphtalate (Behrens \etal 1996) and
2-methyl-2,4-pentanediol (Christensen and Olsen 1995). The
continuous line shows the model, the dashed line the adapted model
(see text).}
\end{figure}

One can do a selfconsistent calculation of $f(V)$ for a given
$f_0(V)$ by inserting the expression for $G(\tau_V)$ in terms of
$f(V)$ into eq. (\ref{f}). This leads to an integral equation for
$f(V)$ with the approximate solution (Buchenau 2002)
\begin{equation}\label{fin}
f(V)=\frac{f_0(V)\exp(-\tau_V/\tau_M)}{\left[3\int_0^\infty\exp(-\tau_V/\tau_v)\exp(-\tau_V/\tau_M)
f_0(v)dv\right]^{2/3}},
\end{equation}
where the Maxwell time is again given by the 1/3-rule, eq.
(\ref{norm}).

The denominator of eq.(\ref{fin}) tends toward zero as $V$
approaches the Maxwell barrier. Therefore the breakdown of the
shear modulus occurs in a rather dramatic way. The relaxing
entities at this critical barrier value receive a strong
enhancement, to such an extent that one is tempted to assume a
separate $\alpha$-process which has nothing to do with the
secondary relaxations. In fact, this more or less unconscious
assumption underlies most of the present attempts to understand
the glass transition (Ediger \etal 1996). The above mean-field
treatment shows such an assumption to be unnecessary; what one
sees at the glass transition are simple Arrhenius relaxations of
no particularly large number density, blown up to impressive size
by the small denominator of eq. (\ref{fin}).

Once $f(V)$ is known, one can determine the frequency dependence
of the complex shear modulus at the frequency $\omega$ from the
two relations
\begin{equation}
G'(\omega)=G\int_0^\infty f(V)
\frac{\omega^2\tau_V^2dV}{1+\omega^2\tau_V^2}\label{Gp}
\end{equation}
and
\begin{equation}
G''(\omega) = G\int_0^\infty f(V) \frac{\omega\tau_VdV}
{1+\omega^2\tau_V^2}\label{Gpp}.
\end{equation}

One must expect $f_0(V)$ and, consequently, $f(V)$ to be a
different function for each glass former. In many of the cases,
however, one should be able to approximate $f_0(V)$ by the
constant value $f_0(V_M)$ for barriers close to $V_M$. Such a
choice is also suggested by the appearance of a constant loss term
(Kudlik \etal 1999; Ngai 2000; Sokolov \etal 2001) in many glass
formers just below $T_g$. A constant loss term corresponds to a
constant $f(V)$. In the model with a barrier-independent $f_0(V)$,
$f(V)$ decreases and approaches $f_0(V)$ as  one goes from the
Maxwell barrier to lower barriers. For this generic case, one can
calculate $G'$ and $G''$ at the breakdown of the shear modulus and
compare the result to measured data.

There are some good mechanical shear measurements over many
decades in frequency at $T_g$ (Christensen and Olsen 1994;
Christensen and Olsen 1995; Behrens \etal 1996; Donth \etal 1996).
Fig. 3 shows data for $G''(\omega)$, normalized to the peak
maximum. The continuous line is calculated from the model assuming
$f_0(V)=const$ and using eqs. (\ref{fin}) and (\ref{Gpp}). As can
be seen from Fig. 3, the fit is only good for the two polymers
polystyrene and poly(vinyl acetate); the two molecular glass
formers show a much stronger decay of $G''$ towards the high
frequency end. For these, the model gives a peak in the imaginary
part of the shear modulus which is too small and too broad; the
real breakdown of the shear modulus is more dramatic than the mean
field calculation. A better fit requires an increase of $f_0(V)$
at the Maxwell barrier (the dashed line in Fig. 3). One can
rationalize this increase; it might be necessary to reach a large
peak in $f_0(V)$ before the shear modulus breaks down.

Dielectric measurements (Kudlik \etal 1999) provide much more
accurate peak shapes than mechanical ones. However, these data are
usually presented as real and imaginary part of the dielectric
constant, which is essentially a susceptibility. The model
discussed here calculates moduli. These should be identical with
the dielectric ones as long as the weighting of the relaxation
centers according to the electric dipole moments corresponds to
the one of the mechanical shear coupling constants. A comparison
requires a conversion of the dielectric constants to dielectric
moduli (Dyre 1991), which changes the peak shape considerably.
Christensen and Olsen (1994) show three examples where dielectric
moduli data give exactly the same peak shape as mechanical shear
modulus data.

\section{Conclusions}

There begins to be a general agreement that the mode coupling
theory of the glass transition (G\"otze and Sj\"ogren 1992) is the
proper microscopic theory for the separation of the structural
relaxation time from the microscopic picosecond time scale in
undercooled liquids. However, the structural relaxation time of
real liquids does not diverge at the critical temperature $T_c$ of
the theory. It rather reaches a a typical value between a
nanosecond and a microsecond (Sokolov 1998), indicating a
crossover to a different flow mechanism.

Numerical simulations (Schr\o der \etal 2000; Angelani \etal 2000;
Broderix \etal 2000; but see also Doliwa and Heuer 2002)
corroborate the old conjecture (Bengtzelius \etal 1984; Angell
1988) that at $T_c$ the system leaves the saddle points of the
energy landscape to settle in the valleys.

At still lower temperatures, one can use the approximation of
harmonic energy minima, with only occasional thermally activated
jumps in between. This is the starting point of a second
microscopic theory (M$\acute{e}$zard and Parisi 1996, 1999) for
the entropy of the undercooled liquid, which allows to calculate
the puzzling heat capacity at the calorimetric glass transition
from the interatomic potentials. These two theories provide some
microscopic insight (though one could wish for theories which are
easier to handle).

The breakdown of the shear modulus at the calorimetric glass
transition has as yet no microscopic explanation. One can show
that it cannot be treated in terms of independent double-well
relaxation centers (Pollak and Pike 1972; Gilroy and Phillips
1981), because the interaction becomes dominant at this breakdown,
reducing the number of relaxation centers needed for the breakdown
by a factor of three (the 1/3-rule). A mean-field treatment
(Buchenau 2002) of the interaction allows to calculate the shear
response at the breakdown. The comparison to measured data shows
agreement in some glass formers, but a more pronounced breakdown
in others. One can rationalize this finding by postulating a rise
of the barrier density function $f_0(V)$ at the Maxwell barrier
for the latter ones.

A disappointing feature of the relaxation center picture is that
one needs a whole temperature-dependent function, the barrier
density function $f_0(V)$, to describe the dynamics around $T_g$.
It is not enough to classify glass formers as type A or B (Kudlik
\etal 1999), depending on whether $f_0(V)$ shows a strong peak
(Johari-Goldstein peak) below $V_M$ or not. On the other hand,
this is a logical consequence of the validity of the energy
landscape idea, because the energy landscape is different in
different glass formers. If all relaxation centers up to the
Maxwell barrier contribute to the flow process, then the
description of its temperature and frequency dependence cannot be
done by a single parameter. Below $T_c$, each glass former
develops its own identity, a conclusion supported by experiment
(Stickel \etal 1995, 1996; Hansen \etal 1997).

\ack An enlightening talk on the current theories of the glass
transition by R. Schilling at the Heraeus Summer School 2002 in
Chemnitz was very helpful. Discussions with W. G\"otze, A. Heuer,
R. Richert and H. R. Schober are gratefully acknowledged.

\section*{References}
\begin{harvard}
\item[] Adam G and Gibbs J H 1965 {\it J. Chem. Phys.} {\bf 43} 139
\item[] Angelani L, Di Leonardo R, Ruocco G, Scala A and Sciortino F 2000
{\it Phys. Rev. Lett.} {\bf 85} 5356
\item[] Angell C A 1988 {\it J. Phys. Chem. Solids} {\bf 49} 863
\item[] Angell C A 1995 {\it Science} {\bf 267} 1924
\item[] Behrens C F, Christiansen T G, Christensen T, Dyre J C and
Olsen N B {\it Phys. Rev. Lett.} {\bf 76} 1553 (1996)
\item[] Bengtzelius U, G\"otze W and Sj\"olander A 1984 {\it J.
Phys. C} {\bf 17} 5915
\item[] B\"ohmer R, Ngai K L, Angell C A and
Plazek D J 1993 {\it J. Chem. Phys.} {\bf 99} 4201
\item[] Broderix K, Bhattacharya K K, Cavagna A, Zippelius A and Giardina I 2000
{\it Phys. Rev. Lett.} {\bf 85} 5360
\item[] Buchenau U 2001 {\it Phys. Rev. B} {\bf 63} 104203
\item[] Buchenau U, Wischnewski A, Zorn R
and Hadjichristides N 2002 {\it Phil. Mag. B} {\bf 82} 209
\item[] Buchenau U 2002 A mean-field model of the glass transition {\it preprint} cond-mat/0202036
\item[] Caliskan G, Kisliuk A, Novikov V  N and Sokolov A P 2002 {\it J. Chem. Phys.} {\bf
114} 10189
\item[] Christensen T and Olsen N B 1994 {\it J. Non-Cryst. Solids} {\bf 172-174} 357
\item[] Christensen T and Olsen N B 1995 {\it Rev. Sci. Instrum.} {\bf 66} 5019
\item[] Doliwa B and Heuer A 2002 Energy barriers and activated dynamics in a supercooled Lennard-Jones
liquid {\it preprint} cond-mat/0209139
\item[] Donth E, Beiner M, Reissig S, Korus J, Garwe F, Vieweg S, Kahle S, Hempel E and
Schr\"oter K 1996 {\it Macromolecules} {\bf 29} 6589
\item[] Dyre J C 1991 {\it J. Non-Cryst. Solids} {\bf 135} 219
\item[] Ediger M D, Angell C A and
Nagel S R 1996 {\it J. Phys. Chem.} {\bf 100} 13200
\item[] Ferry D J 1980 {\it Viscoelastic properties of
polymers} (New York: 3rd ed., John Wiley)
\item[] Frick B, Zorn R, Richter D and Farago B 1991 {\it J.
Non-Cryst. Solids} {\bf 131} 169
\item[] Gilroy K S and Phillips W A 1981
{\it Phil. Mag. B} {\bf 43} 735
\item[] Franosch T, G\"otze W, Mayr M R and Singh A P
1998 {\it J. Noncryst. Solids} {\bf 235-237} 71

\item[] G\"otze W and Sj\"ogren L 1992 {\it Rep. Prog. Phys.} {\bf
55} 241
\item[] G\"otze W 1999 {\it J. Phys: Condens. Matter} {\bf
11} A1-A45
\item[] Goldstein M 1968 {\it J. Chem. Phys.} {\bf 51}, 3728
\item[] Hansen C, Stickel F, Berger T, Richert R and Fischer E. W.
1997 {\it J. Chem. Phys.} {\bf 107} 1086
\item[] Heuer A and Silbey R J 1996 {\it Phys. Rev. B} {\bf 53} 609
\item[] J\"ackle J 1986 {\it Rep. Prog. Phys.} {\bf 49} 171
\item[] Johari G P and Goldstein M 1970 {\it J. Chem. Phys.} {\bf 53} 2372
\item[] Johari G P and Goldstein M 1971 {\it J. Chem. Phys.} {\bf 55} 4245
\item[] Johari G P 1986 {\it Phys. Rev. B} {\bf 33} 7201
\item[] Kammerer S, Kob W and Schilling R 1998a {\it Phys. Rev. E} {\bf 58} 2131
\item[] Kammerer S, Kob W and Schilling R 1998b {\it Phys. Rev. E} {\bf 58} 2141
\item[] Knaak W, Mezei F and Farago B 1988 {\it Europhys. Lett.} {\bf 7} 529
\item[] Kob W and Andersen H C 1995a {\it Phys. Rev. E} {\bf 51} 4626
\item[] Kob W and Andersen H C 1995b {\it Phys. Rev. E} {\bf 52} 4134
\item[] Kudlik A, Benkhof S, Blochowicz T, Tschirwitz C and R\"ossler E 1999
{\it J.Molec.Structure} {\bf 479} 201
\item[] Li G, Du W M, Chen X K, Cummins H Z and Tao N 1992 {\it
Phys. Rev. A} {\bf 45} 3867
\item[] M$\acute{e}$zard M and Parisi G 1996 {\it J. Phys.} {\bf A29} 6515
\item[] M$\acute{e}$zard M and Parisi G 1999 {\it J. Chem. Phys.} {\bf 111} 1076
\item[] Nauroth M and Kob W 1997 {\it Phys. Rev. E} {\bf 55} 657
\item[] Ngai K L 2000 {\it J. Non-Cryst. Solids} {\bf 275} 7
\item[] Parshin D. A. 1994 {\it Phys. Solid State} {\bf 36} 991
\item[] Phillips W A 1981 (ed.), {\it Amorphous
Solids: Low temperature properties}, (Berlin: Springer)
\item[] Pollak M and Pike G E 1972 {\it Phys. Rev. Lett.} {\bf
28} 1449
\item[] Pusey P N and van Megen W 1986 {\it Nature}
{\bf 320} 340
\item[] Richert R. 2002 {\it J. Phys.: Condens. Matter} {\bf 14}
R703-R738
\item[] Schober H R and Oligschleger C 1996 {\it Phys. Rev. B} {\bf 53} 11469
\item[] Schr\o der T B, Sastry S, Dyre J C and Glotzer S C 2000 {\it
J. Chem. Phys.} {\bf 112} 9834
\item[] Sokolov A P 1998 {\it J. Non-Crystalline Solids} {\bf 235-237} 190
\item[] Sokolov A P, Novikov V N, Kisliuk A and Ngai K L 2001 {\it Phys. Rev. B} {\bf
63} 172204
\item[] Stickel F, Fischer E. W. and Richert R 1995
{\it J. Chem. Phys.} {\bf 102} 6251
\item[] Stickel F, Fischer E. W. and Richert R 1996
{\it J. Chem. Phys.} {\bf 104} 2043
\item[] Stillinger F H 1995 {\it Science} {\bf 267} 1935
\item[] Surovtsev N V, Wiedersich J, Novikov V N, R\"ossler E and Sokolov
A P 1998 {\it Phys. Rev. B} {\bf 58} 14888
\item[] Taborek P, Kleiman R N and Bishop D J 1986 {\it Phys.
Rev. B} {\bf 34} 1835
\item[] van Megen W and Underwood S M 1994 {\it Phys.
Rev. E} {\bf 49} 4206
\item[] Wiedersich J, Surovtsev N V and R\"ossler E 2000a
{\it J. Chem. Phys.} {\bf 113} 1143
\item[] Wiedersich J, Adichtchev S V and R\"ossler E 2000b
{\it Phys. Rev. Lett.} {\bf 84} 2718
\item[] Wuttke J, Kiebel M, Bartsch E, Fujara F, Petry W and
Sillescu H 1993 {\it Z. Phys. B: Cond. Matt.} {\bf 91} 357
\end{harvard}

\end{document}